\documentclass[twocolumn,prc,aps, nofootinbib]{revtex4}
\usepackage{graphicx}
\usepackage[section]{placeins}

 \def\Pom{{ I\!\!P}}
 
 \def\gsim{\mathrel{\rlap{\lower4pt\hbox{\hskip1pt$\sim$}}
 \raise1pt\hbox{$>$}}}

 \newcommand\la{\langle}
 \newcommand\ra{\rangle}
 \newcommand\beq{\begin{equation}}
 
 \newcommand\eeq{\end{equation}}
 \newcommand\beqn{\begin{eqnarray}}
 \newcommand\eeqn{\end{eqnarray}}
\def\mb{\,\mbox{mb}}
\def\fm{\,\mbox{fm}}
\def\GeV{\,\mbox{GeV}}
\def\TeV{\,\mbox{TeV}}
\def\lsim{\mathrel{\rlap{\lower4pt\hbox{\hskip1pt$\sim$}}
    \raise1pt\hbox{$<$}}}         
\def\gsim{\mathrel{\rlap{\lower4pt\hbox{\hskip1pt$\sim$}}
    \raise1pt\hbox{$>$}}}         
\def\Re{\,\mbox{Re}\,}
\def\Im{\,\mbox{Im}\,}
\def\mb{\,\mbox{mb}}
\def\fm{\,\mbox{fm}}
\def\GeV{\,\mbox{GeV}}

\def\doublespace{\def\baselinestretch{1.6}\large\normalsize}
\def\normalspace{\def\baselinestretch{1.0}\normalsize}
\def\Caption#1{\normalspace
  \begin{quotation}\caption{\sl #1}\end{quotation}
  \doublespace}
\def\beq{\begin{equation}}
\def\eeq{\end{equation}}

\def\beqy{\begin{eqnarray}}
\def\eeqy{\end{eqnarray}}

\begin{document}

\title{\bf Two-scale hadronic structure and elastic pp scattering:\\
predicted and measured}

\author{B. Z. Kopeliovich$^1$}
\author{ I. K. Potashnikova$^1$}
\author{B.~Povh$^{2}$}
\affiliation{\centerline{$^1$Departamento de F\'{\i}sica,
Universidad T\'ecnica Federico Santa Mar\'{\i}a; and}
Instituto de estudios avanzados en ciencias en ingenier'a; and\\
Centro Cient\'ifico-Tecnol\'ogico de Valpara\'iso;\\
Casilla 110-V, Valpara\'iso, Chile
\\
{$^{2}$ Max-Planck-Institut f\"ur Kernphysik, Postfach 103980, 69029 Heidelberg, Germany
}}

\begin{abstract}
We update the comparison with experiment of the dynamical model of high-energy hadron interactions based on the two scale structure of hadrons. 
All predictions made over decade ago are confirmed with a high precision by the TOTEM experiment at LHC.
\end{abstract}



\maketitle

\section{Introduction}

The results of TOTEM experiment published recently \cite{totem11,totem12}
on small angle elastic $pp$ scattering at $\sqrt{s}=7\TeV$ have a non-precedential
accuracy and suggest a stringent test of contemporary models. In this note we confront with data the predictions of the dynamical model for the elastic amplitude, which was proposed in \cite{k3p-1,k3p-2}. The model is based on the presence of two scales in hadronic structure, a soft one, of the order of the confinement radius $R_c\sim1/\Lambda_{QCD}\approx 1\fm$, and a semi-hard scale $r_0\approx 0.3\fm$, which originates from the non-perturbative interactions of gluons. The model allows to predict in a least parameter-dependent way the energy dependence of the total and elastic differential cross sections.  
\begin{figure}[h]
 \includegraphics[height=7.5cm]{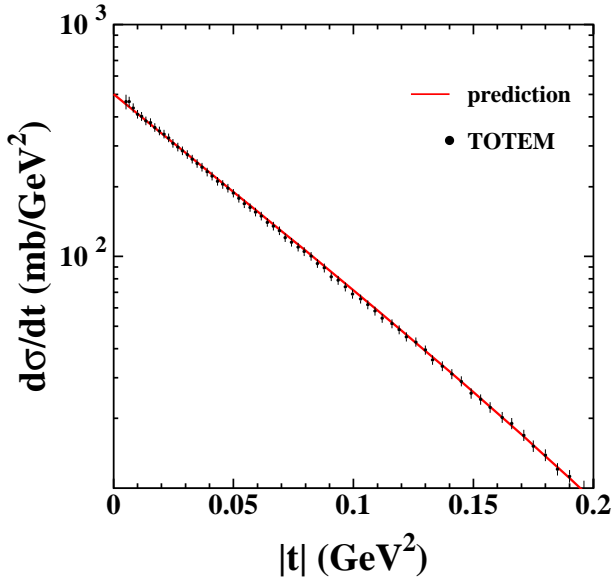}
\caption{ \label{fig:ds-dt-k3p} (Color online)
The differential elastic cross section predicted in \cite{k3p-1,k3p-2}  and measured by TOTEM \cite{totem11,totem12}.
}
 \end{figure}
The model prediction for the differential cross section of $pp$ elastic scattering at $\sqrt{s}=7\TeV$ is compared with the TOTEM data in Fig.~\ref{fig:ds-dt-k3p} at $|t|<0.2\GeV^2$, the interval which was used in \cite{totem12} for determination of the $t$-slope.
The accuracy of data, and the agreement with predictions made 12 year prior the measurements look astonishing, both for the absolute value and the slope. Below we briefly overview the model and perform a detailed comparison with different observables.

\section{The two-scale model}

Data on diffractive excitation of a proton $pp\to pX$ to large invariant mass $M_X$, dominated by diffractive gluon radiation, 
demonstrate amazingly small cross section, known in Regge phenomenology as smallness of the triple-pomeron term. Its smallness can be explained by an increased
mean transverse momentum of the gluons, $k_g\sim0.7\GeV$, i.e. by a small size of the gluonic spots in the hadron. There are many other experimental evidences \cite{spots} for presence such a small parameter characterizing the hadron structure.
The presence of such a semi-hard scale is also supported by the nonperturbative QCD models. A small scale similar to $r_0$ characterizing the nonperturbative interaction of gluons, also emerges from the lattice calculations \cite{lattice}, and from the instanton liquid model \cite{shuryak1,shuryak2}.

The light-cone distribution function of a quark-gluon Fock state in the limit of small fractional momentum of the gluon was derived in \cite{kst2} in the form,
\beq
\Psi_{qg}(\vec r)=
-\,\frac{2\,i}{\pi}\,\sqrt{\alpha_s\over3}\,\,
\frac{\vec e_g^{\,*}\cdot\vec r}{r^2}\,
\exp\left(-\,\frac{r^2}{2\,r_0^2}\right)\ ,
\label{100}
\eeq
where $r$ is the transverse quark-gluon separation;
$r_0=0.3\fm$ was adjusted to explain the observed weakness of large mass diffraction;
$\vec e$ is the gluon polarization vector. 

Smallness of $r_0$ justifies an attempt to calculate gluon radiation perturbatively with the distribution function (\ref{100}).
Moreover, for a dipole, as small as $r_0$, one can rely on the quadratic form of the dipole-nucleon cross section, $\sigma_{\bar qq}(r)=C\,r^2$.
The cross section of radiation of $n$ gluons, derived in Refs.~3-4, reads,
 \beq
\sigma_n = \frac{1}{n!}\,
\left[\frac{4\,\alpha_s}{3\,\pi}\,\, 
{\rm ln}\left({s\over s_0}\right)\right]^n\,
\frac{9}{4}\,C\,r_0^2\ . 
\label{120} 
\eeq
One should be cautious, however, applying this formula at $n=0$.
Even if the intrinsic transverse momenta of gluons are very large, so that the radiation cross section vanishes, the nonperturbative cross section of interaction
of two large colliding dipoles remains large. The value of energy independent term $\sigma_0$
is controlled by nonperturbative effects and hardly can be reliably evaluated with the available theoretical tools. This is why it was treated in \cite{k3p-1,k3p-2} as a fitted parameter, in fact the only parameter of the model.

Thus, summing up the powers of logarithms in (\ref{120}) we arrive at
the total cross section, which has the form,
 \beq 
\sigma_{tot}=\sum\limits_{n=0}\sigma_n
= \tilde\sigma_0 +
\frac{27}{4}\,C\,r_0^2\,\, 
\left({s\over s_0}\right)^{\Delta}\ , 
\label{140} 
\eeq 
 where $\tilde\sigma_0=\sigma_0-(27/4)C\,r_0^2$;
 $\Delta=4\alpha_s/(3\pi)$; and $s_0=30\GeV^2$.
 
 The peculiar structure of Eq.~(\ref{140}) bares the general character.
 The cross section contains two terms, one is energy independent, controlled by the poorly known nonperturbative dynamics. It is probably large and dominates the cross section at low energies. The other term has the Regge pole structure, rises with energy and dominates
 the cross section at very high energies. It is very unlikely that the first term might be zero,
 what would need a fine tuning between the nonperturbative and perturbative contributions to exponentiate together, i.e. to cancel each other in $\tilde\sigma_0$.

The second, rising with energy term in (\ref{140}) can be rather well evaluated.
The semi-hard scale $0.3\fm$ is sufficiently small to justify  an attempt to use perturbative QCD calculations. Important for Eqs.~(\ref{120})-(\ref{140}) is
the value of $\alpha_s$ at this scale. It was evaluated in Refs.~3-4 in different ways. Assuming that the semi-hard scale $r_0$ characterizes the transition from the perturbative regime to the regime of spontaneous chiral symmetry breaking \cite{gribov,ewerz}, one can rely on the critical value $\alpha_c = (3\pi/4)(1-\sqrt{2/3})$. Another estimate corresponds to the mean value of $\alpha_s$ averaged over the gluon radiation spectrum, as was done in \cite{k3p-2}. Both methods converge
at a value $\alpha_s\approx 0.4$, which was used in Refs.~3-4. Then one arrives at
  \beq 
 \Delta=\frac{4\,\alpha_s}{3\,\pi}=0.17\ . 
\label{160} 
\eeq 
This value is about twice as large as $\epsilon=0.08$ fitted to describe the energy dependence of the total hadronic cross sections, parametrized as $s^\epsilon$.
However, presence in (\ref{140}) of the large constant term makes the rise of the cross section less steep, in good agreement with data \cite{k3p-1,k3p-2}.
The factor $C=2.3$ of Eq.~(\ref{140}) was also calculated in Refs.~3-4 within the two-gluon Born mechanism (since gluon radiation is included explicitly) using the same value of $\alpha_s$ .

The Pomeron partial elastic amplitude corresponding to the cross section Eq.~(\ref{140}),  has the form
\beq
\Im \gamma_{\Pom}(s,b)=
\sum\limits_{n=0}\Im\gamma_n(s,b),
\label{180}
\eeq
where the shape of $\Im\gamma_n(s,b)$ was derived in \cite{k3p-2} basing on the results of \cite{huf-povh},
 \beq
{\rm Im}\,\gamma_n(s,b)=
\frac{\sigma^{hN}_n(s)}{8\,\pi\,B_n}\,
y^3\,K_3(y).
\label{200}
\eeq
 Here $y^2=(4b^2/B_n)^3$, $K_3(y)$ is the third order modified
Bessel function. The slope corresponding to radiation of $n$ gluons rises
with $n$ due to the Brownian motion of gluons in the transverse plane,
 \beq
B_n = {2\over3}\,\left\la r_{ch}^2\right\ra +
\frac{n\,r_0^2}{2}\ .
\label{220}
\eeq

Notice that the mean number of radiated gluons according to (\ref{120}) rises with energy as
\beq
\la n\ra=\frac{4\alpha_s}{3\pi}\,\ln(s/s_0).
\label{240}
\eeq
Comparing with Eq.~(\ref{220}) we conclude that the effective slope $\alpha_{\Pom}^\prime$ of the Pomeron trajectory reads \cite{k3p-2,spots},
\beq
\alpha_{\Pom}^\prime=\frac{\alpha_s}{3\pi}\,r_0^2 = 0.1\GeV^{-2}.
\label{260}
\eeq
This value, at first glance, is quite smaller than the known $\alpha_{\Pom}^\prime=0.25\GeV^{-2}$, which is required to describe the observed shrinkage of the diffractive cone. As we will see further, the situation is more involved and more interesting, and the value Eq.~(\ref{260}) well agrees with data.

The the total cross section Eq.~(\ref{140}) and the partial amplitude Eq.~(\ref{180}) rising as a power of energy will eventually break the unitarity bound. Therefore one should apply a procedure usually called unitarization. This is probably the most uncertain point
of every dynamical model for the elastic amplitude. In Refs.~3-4 the preference was given to the popular quasi-eikonal model of unitarization \cite{kaidalov}, which modifies the partial elastic amplitude as,
 \beq
\Im\Gamma_{\Pom}(s,b)=
\frac{1}{K(s)}\,\left[
1 - e^{-K(s)\,{\rm Im}\,\gamma_{\Pom}(s,b)}
\right],
\label{280}
\eeq
where $K(s)=1+\sigma_{sd}/\sigma_{el}$, and $\sigma_{sd}$ is the single diffractive cross section (see more details in \cite{k3p-2}) . 
Here at least the first two terms (most important) of expansion of the exponential in (\ref{280}) are correct, as follows from the AGK cutting rules \cite{agk}.

\subsection{Total cross section}

\begin{figure}[h]
 \includegraphics[height=7cm]{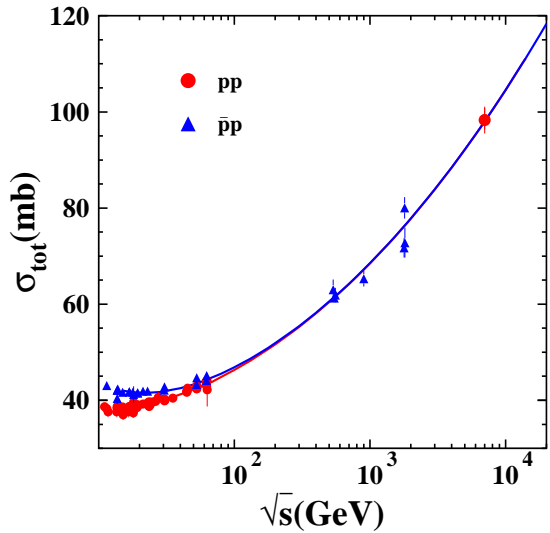}
\caption{ \label{fig:stot} (Color online)
Total cross section predicted in \cite{k3p-1,k3p-2} and measured by TOTEM
\cite{totem11,totem12}.
}
 \end{figure}

Eq.~(\ref{280}) is the final result of the model, which can be employed calculating different observables. The first is the total cross section,
\beq
\sigma^{pp}_{tot}(s)=2\int d^2b\,\Im\Gamma_{\Pom}(s,b).
\label{300}
\eeq
All the ingredients of the partial amplitude are known, as was explain above, except one parameter $\tilde\sigma_0$ in (\ref{140}) to be fitted to data. Since it is independent of energy, it was fixed in \cite{k3p-1,k3p-2} at $\tilde\sigma_0=39.7\mb$ by comparison with a single experimental point \cite{cdf} at $\sqrt{s}=546\GeV$.
The energy dependence is then predicted with no further adjustment as is demonstrated in Fig.~\ref{fig:stot}.
The value  $\sigma^{pp}_{tot}=98\mb$ predicted at $7\TeV$ exceedingly well agrees with the result of the TOTEM experiment \cite{totem12}, $\sigma^{pp}_{tot}=98.58\pm2.23\mb$.

Notice that the contribution of Reggeons was added in \cite{k3p-1,k3p-2} to discriminate between $pp$ and $\bar pp$ collisions. That contribution was not predicted, but fitted, however it has little to do with the high energy range under discussion.

\subsection{Real part of the elastic amplitude}

We rely on the derivative analyticity relation \cite{gribov-migdal,bronzan,pavel1,pavel2} between the real and imaginary parts of the elastic amplitude,
\beq
\Re f_{el}(s,t) = {\pi\over2}\,\frac{\partial}{\partial\ln s}\,\Im f_{el}(s,t),
\label{320}
\eeq
where
\beq
\Im f_{el}(s,t)=
\int d^2b\,
e^{i\vec b\cdot\vec q}\,
\Im\Gamma_{\Pom}(s,b),
\label{340}
\eeq
and $t=-\vec q\,^2$

\begin{figure}[h]
\includegraphics[height=4.2cm]{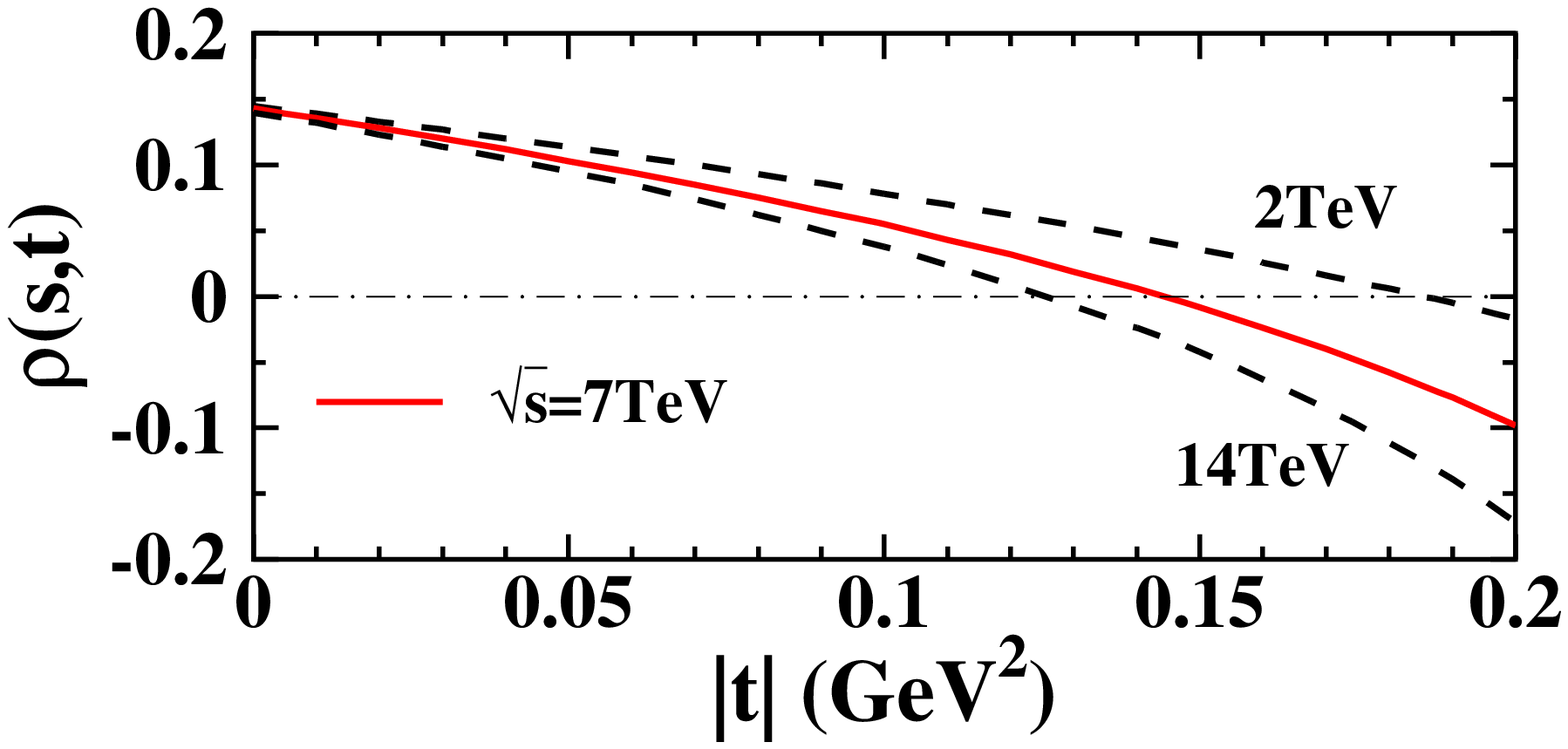}
 \includegraphics[height=4.2cm]{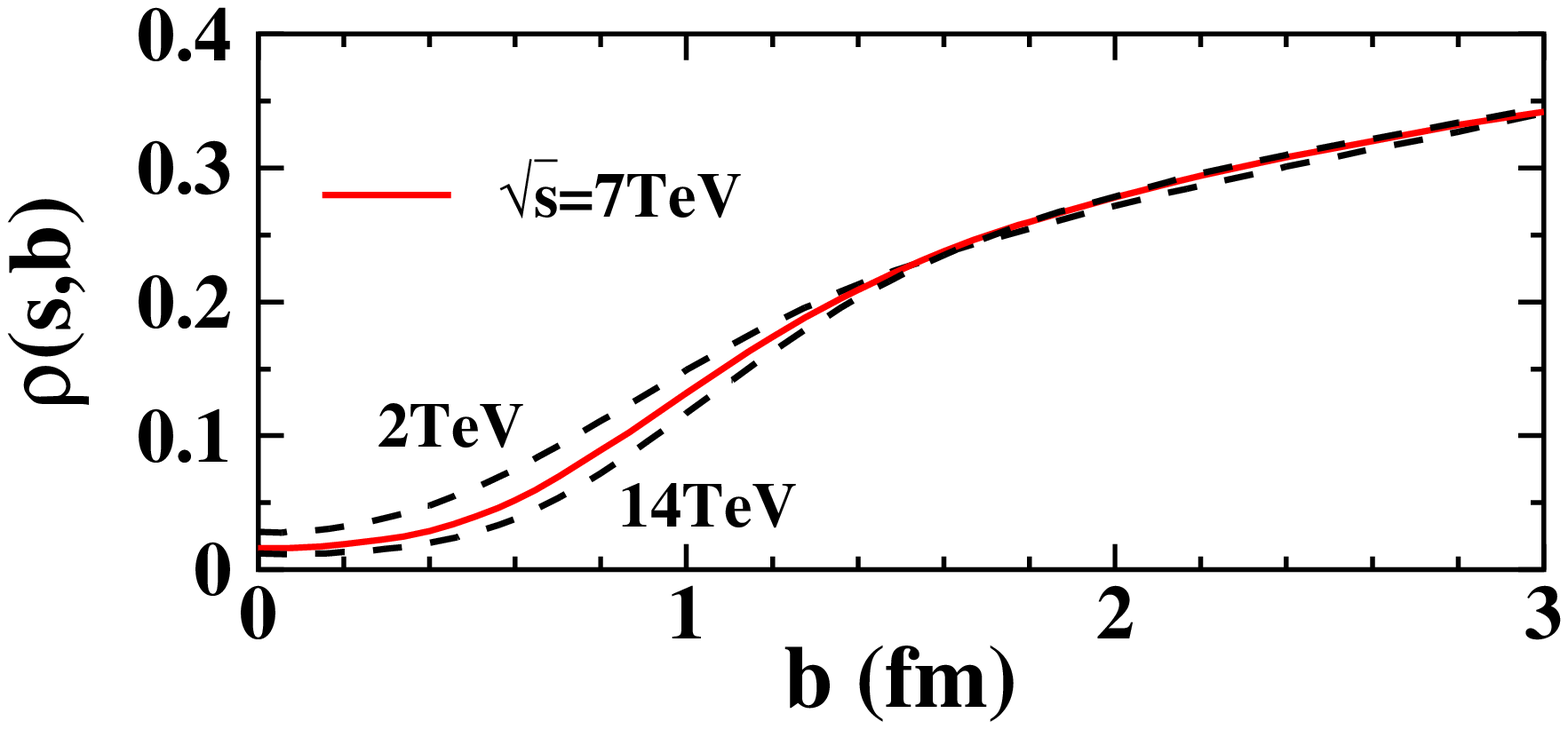}
\caption{ \label{fig:rho} (Color online)
Ratio of real to imaginary parts of elastic $pp$ amplitude as function of $t$
(upper panel, Eqs.~(\ref{320})-(\ref{340})), and as function of impact parameter (bottom panel, Eq.~(\ref{360})) at $\sqrt{s}=2,\ 7$ and $14\TeV$.
}
 \end{figure}

Correspondingly, one can evaluate the ratio of real-to-imaginary parts of the amplitude, $\rho(s,t)=\Re f_{el}(s,t)/\Im f_{el}(s,t)$,  as function of $t$. The results at $\sqrt{s}=2,\ 7$ and $14\TeV$ are depicted in the upper panel of Fig.~\ref{fig:rho}.
Notice that the phase of the elastic amplitude can be measured at $t\to0$
via the interference with the Coulomb amplitude, and our result $\rho(t=0)=0.143$
well agrees with the fit \cite{compete} to data, $\rho(0)=0.141\pm 0.007$.

It is also instructive to see how the value of $\rho$ depends on impact parameter. 
According to Eq.~(\ref{320}),
\beq
\rho(s,b)=\frac{\pi}{2}\,\Delta_{eff}^{\Pom}(s,b),
\label{360}
\eeq
where the effective Pomeron intercept in impact parameter representation was calculated in Ref.~3-4 as,
\beq
\Delta_{eff}^{\Pom}(s,b)=
\frac{\partial}{\partial\ln s}\ln\left[\Im\Gamma_{\Pom}(s,b)\right].
\label{370}
\eeq
$\rho(s,b)$ is plotted in the bottom panel of Fig.~\ref{fig:rho} at $\sqrt{s}=2,\ 7$ and $14\GeV$ .
Naturally $\rho(b=0)$ is vanishing upon approaching the black disk limit at
high energies.

\subsection{Elastic cross section}

The elastic differential cross section,
\beq
\frac{d\sigma_{el}}{dt}=\frac{1}{4\pi}\left| f_{el}(s,t)\right|^2,
\label{380}
\eeq
calculated at $\sqrt{s}=7\TeV$ is plotted in Fig.~\ref{fig:ds-dt-k3p}
in comparison with TOTEM data \cite{totem12}. The agreement is excellent.
Notice that the correction from inclusion of the real part of the amplitude is quite small,
less than $2\%$.

\begin{figure}[h]
 \includegraphics[height=7cm]{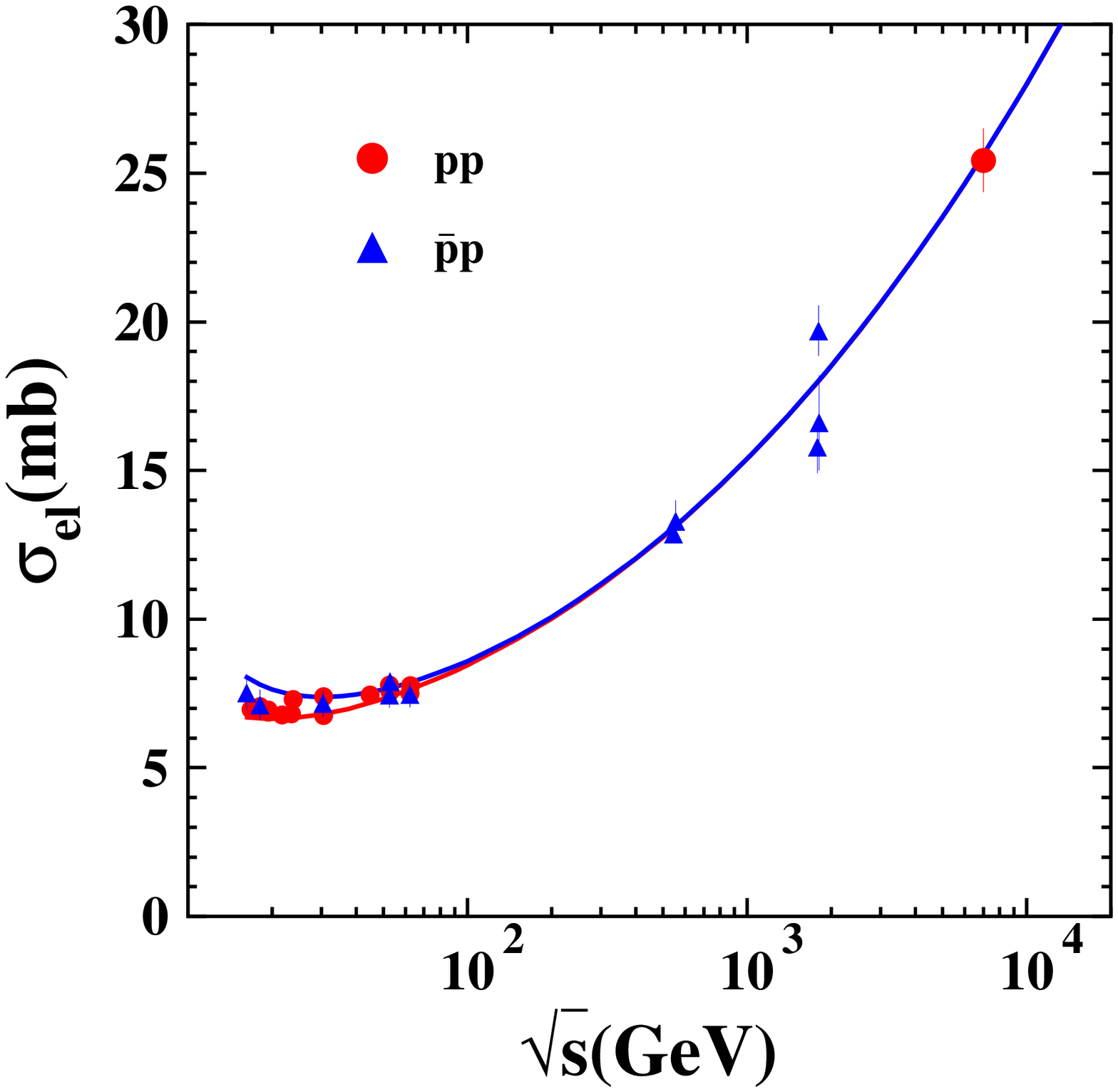}
\caption{ \label{fig:sel} (Color online)
Elasic cross section predicted in \cite{k3p-1,k3p-2} and measured by TOTEM
\cite{totem11,totem12}.
}
 \end{figure}

Integrating this cross section over $t$ we arrive at the elastic cross section,
\beq
\sigma^{pp}_{el}(s)=\int d^2b\,\left[\Im\Gamma_{\Pom}(s,b)\right]^2
\left[1+\rho^2(s,b)\right],
\label{390}
\eeq
which is plotted vs energy in Fig.~\ref{fig:sel}.
 Again the predicted cross section is well confirmed by the TOTEM measurement 
\cite{totem11,totem12}. This is not a surprise, since the differential cross section
shown in Fig.~\ref{fig:ds-dt-k3p} well agrees with the data in the interval of $t$, which provides the dominant contribution to the integrated elastic cross section.

\begin{figure}[h]
 \includegraphics[height=7cm]{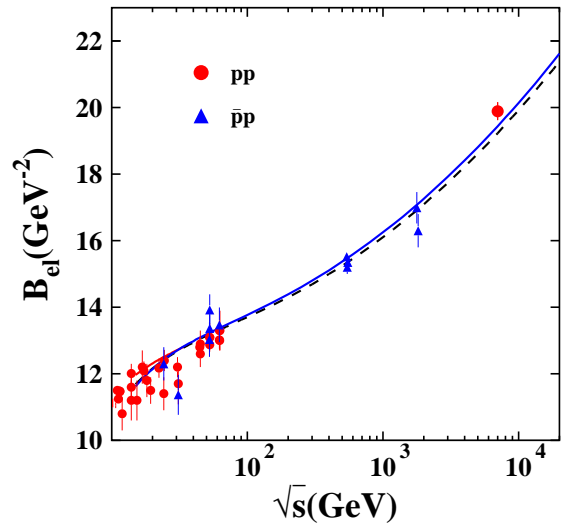}
\caption{ \label{fig:Bel} (Color online)
Elastic slope predicted at $t=0$ in \cite{k3p-1,k3p-2} and measured by TOTEM
\cite{totem11,totem12}. Dashed and solid curves show $B_{el}$, calculated excluding or including the real part of the amplitude, respectively.
}
 \end{figure}

The $t$-slope of the differential cross section at $t=0$ was predicted in \cite{k3p-1,k3p-2}
in a parameter-free way, after the only parameter $\tilde\sigma_0$ was fixed 
by data on $\sigma^{pp}_{tot}$.
\beq
\left.B_{el}^{\rm (Im)}(s,t)\right|_{t=0}=\frac{1}{\sigma^{pp}_{tot}}
\int d^2b\,b^2\,\Im\Gamma_{\Pom}(s,b).
\label{400}
\eeq
Adding the upper-script $(\Im)$ we emphasize that the real part of the amplitude was neglected. The predicted $B_{el}^{\rm (Im)}$ is plotted in Fig.~\ref{fig:Bel} by dashed curve. 

Although the result of Eq.~(\ref{400}) $B_{el}^{\rm (Im)}=19.23\GeV^{-2}$ agrees rather well with the TOTEM measurement $B_{el}(\sqrt{s}=7\TeV)=19.89\pm0.27$,
we nevertheless perform also calculations including the small correction from the real part.
\beq
\left. B_{el}(s,t)\right|_{t=0}=\left.\frac{B_{el}^{\rm (Im)}(s,t)+
\rho(t)\,B_{el}^{\rm (Re)}(s,t)}{1+\rho^2(t)}
\right|_{t=0}\!\!\!,
\label{420}
\eeq
where
\beq
\left.B_{el}^{\rm (Re)}(s,t)\right|_{t=0}=
\frac{1}{\sigma^{pp}_{tot}}
\int d^2b\,b^2\,\Re\Gamma_{\Pom}(s,b)
\label{440}
\eeq
The full slope Eq.~(\ref{420}) is plotted by solid curve in Fig.~\ref{fig:Bel}.
As expected, the difference between the dashed and solid curves is very small.

\begin{figure}[h]
 \includegraphics[width=5cm]{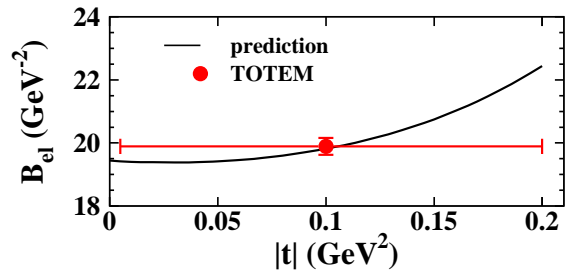}
\caption{ \label{fig:Bel-t} (Color online)
Variation of the predicted \cite{k3p-1,k3p-2} $B_{el}$ with $t$ and measured by TOTEM
within the interval $0.005<|t|<0.2\GeV^2$ \cite{totem11,totem12}.
}
 \end{figure}

Although the predicted and measured values of the elastic slope are close, as is depicted in
Fig.~\ref{fig:Bel}, in fact it is better than looks. We calculated the slope at $t=0$,
because in most previous experiments it was measured at very small $|t|\approx 0.02\GeV^2$.
However, in the TOTEM experiment the slope was determined within the rather large interval
$0.005<|t|<0.2\GeV^2$. As is shown in Fig.~\ref{fig:Bel-t} the calculated slope slightly rises with $|t|$ and hits exactly the measured value of $B_{el}$ in the middle of the used $t$-interval.
Although the data agree with a $t$-independent slope, the error bar would be larger if
only points with very small $|t|\sim 0.02\GeV^2$ were used for the fit.

The successful prediction of $B_{el}(s)\approx B_0+2\alpha^\prime\ln(s/s_0)$ brings us back to the problem raised by
the Eq.~(\ref{260}): how did we manage to get the correct energy dependence of the slope $B_{el}(s)$ with such a small value of
$\alpha_{\Pom}^\prime$?
The answer is obvious \cite{spots}, the observed steep rise with energy of $B^{pp}_{el}(s)$ is a result of saturation. Namely, while at small impact parameters the
partial amplitude nearly saturates the unitarity bound, $\Im\Gamma(s,b)\to1$, 
it is still small for peripheral collisions, and keeps rising with energy. As a result,
the mean value $\la b^2\ra=2B_{el}$ would increase with energy even if the input value of $\alpha_{\Pom}^\prime$ were zero. So the unitarity (absorption) corrections is the main contributor to the observed rise of the elastic $t$-slope with energy.
We can test this conclusion checking with the processes, which have a smaller cross section and are still far from the unitarity bound. In such processes one should observe the genuine 
$\alpha_{\Pom}^\prime$. Indeed, the Pomeron contribution to diffractive
charmonium photoproduction, $\gamma+p\to J/\Psi+p$, should be free either of the secondary Reggeons, or of the absorptive corrections. Indeed, data from the ZEUS experiment \cite{zeus} require a small value of
$\alpha_{\Pom}^\prime=0.115\pm0.018$ in good agreement with Eq.~({\ref{260}).

\section{Summary and outlook}

Here we tested thoroughly the 12-year old predictions of Refs.~3-4 for elastic $pp$ scattering, comparing with different observables measured recently in the TOTEM experiment at LHC and $\sqrt{s}=7\TeV$. The predictions were made within the model based on the two-scale hadronic stricture: the soft and semi-hard scales related either to
the confinement radius, or to the small size of gluonic spots, respectively. The enhanced 
mean transverse momentum of gluons allows to perform parameter-free perturbative calculations for
gluon radiation, which is the source of the rising energy dependence of the cross sections.
The two scales in the hadron structure leads to a specific form of the energy dependent Pomeron amplitude, which consists of two terms, a constant one, related to the poorly known nonperturbative dynamics, and a term steeply, as $s^{0.17}$, rising with energy.
The latter is fully calculated in a parameter-free way, while the former is treated as a parameter, which controls the absolute value of the cross section (but not its energy dependence). 
We observe an excellent agreement between the predictions made in Refs.~3-4 and the TOTEM data.

It is worth noting that our predictions were not a kind of simple extrapolation  of available data to higher energies using some parametrization, as is frequently done in the literature. We calculated all the parameters, except a single one, representing the non-perturbative physics.

Of course our predictions have some theoretical uncertainty, which we believe is related mainly to some freedom in the choice of the value of $\alpha_s=0.4$. The corresponding
effective intercept was estimated in Ref.~4 at $\Delta=0.17\pm0.01$. Keeping the normalization
of $\sigma^{\bar pp}_{tot}$ at $\sqrt{s}=546\GeV$ unchanged we found the uncertainty
of our prediction to be about $6\%$ at the energies of LHC. The good agreement with the TOTEM data might indicate at correctness of the central value of $\Delta$ chosen in Refs.~3-4.

In our calculations here we did not make any hidden modification
to the model of Refs.~3-4, or any adjustment to the TOTEM data,
because we wanted the results to keep the status of {\it predictions}.
This does not mean, however, that there is no room for further improvements in the model.
In particular, the popular quasi-eikonal model for unitarization  is poorly justified.
For this reason, as was emphasized in Refs.~3-4, the model does not pretend for a good description of elastic scattering at large $t$, in particular in the region of the diffractive minimum, what would need a much higher accuracy of the model. Meanwhile, a deeper understanding and much better theoretical tools for calculation of absorptive corrections have been developed over the past decade \cite{kps}, and we plan to
work on that, aiming at the improvement of the description of elastic scattering at larger momentum transfer.

In conclusion, we summarize the predictions displayed in Table~\ref{tab-predictions} for the basic observables measured at $\sqrt{s}=7\TeV$, and expected to be measured, at $8$ and $14\TeV$.
\begin{table}[h]
 \Caption{
 \label{tab-predictions}
 Predictions of Refs.~3-4 for the cross sections and elastic slopes at $\sqrt{s}=7,\ 8$ and $14\TeV$. Two values of the slope correspond to $t=0$ and $t=-0.1\GeV^2$ (in parentheses).}
 \begin{center}
\begin{tabular}{|c|c|c|c|}
\\[-12mm]
 \hline
 \vphantom{\bigg\vert}
   $\sqrt{s}$ 
   & $\sigma^{pp}_{tot}(\mb)$
  & $\sigma^{pp}_{el}(\mb)$
  & $B_{el}(\GeV^{-2})$  \\
\hline &&&\\[-6mm]
$7\TeV$& 98.00 & 25.63 & 19.44 (19.82) \\
\hline &&&\\[-6mm]
$8\TeV$& 100.41 & 26.50 & 19.70 (20.19) \\
\hline &&&\\[-6mm]
$14\TeV$ & 111.07 & 30.39 & 20.84 (21.92) \\
\hline   
\end{tabular}
\end{center}
 \end{table}

\acknowledgments
This work was supported in part by Fondecyt (Chile) grants 1090236 and
1090291. The work of B.Z.K. was also 
supported by the Alliance Program of the Helmholtz Association, contract HA216/EMMI "Extremes of Density and Temperature: Cosmic Matter in the Laboratory".

\end{document}